%% file: main.tex
\documentclass[12pt,a4paper,leqno]{article}

\usepackage{amssymb,exscale}
\usepackage[centertags]{amsmath}
\usepackage{amsthm}
\usepackage{graphicx}

\input{preamble}

\newcommand{\Br}{W}

\begin{document}

\title{A non-Fock fermion toy model}
\author{Boris Tsirelson}
\date{}

\maketitle

\begin{abstract}
Recent progress in mathematical theory of random processes provides us
with non-Fock product systems (continuous tensor products of Hilbert
spaces) used here for constructing a toy model for fermions. Some
state vectors describe infinitely many particles in a finite region;
the particles accumulate to a point. Electric charge can be assigned
to the particles, the total charge being zero. Time dynamics is not
considered yet, only kinematics (a single time instant).
\end{abstract}

\section*{Introduction}
\input{intro}

\section{The idea}
\input{sect1}

\section{The model}
\input{sect2}

\section{Toward more realistic models}
\input{sect3}

\bigskip
\filbreak
\begingroup
{
\small
\begin{sc}
\parindent=0pt\baselineskip=12pt

School of Mathematics, Tel Aviv Univ., Tel Aviv
69978, Israel
\emailwww{tsirel@math.tau.ac.il}
{http://www.math.tau.ac.il/$\sim$tsirel/}
\end{sc}
}
\filbreak

\endgroup

\end{document}

%% file: preamble.tex
\numberwithin{equation}{subsection}

\swapnumbers
\theoremstyle{definition}

\renewcommand{\phi}{\varphi}

\renewcommand{\(}{\bigl(}
\renewcommand{\)}{\bigr)\vphantom{)}}

\newcommand{\sgn}{\operatorname{sgn}}

\newcommand{\Var}{\operatorname{Var}}
\newcommand{\argmax}{\operatorname{arg\,max}}
\newcommand{\argmin}{\operatorname{arg\,min}}
\renewcommand{\Re}{\operatorname{Re}}

\newcommand{\bN}{\mathbf N}

\newcommand{\eps}{\varepsilon}

\newcommand{\F}{\mathcal F}

\newcommand{\la}{\lambda}

\newcommand{\Ex}{\mathbb E}

\newcommand{\R}{\mathbb R}
\newcommand{\C}{\mathbb C}
\newcommand{\Z}{\mathbb Z}

\newcommand{\cE}[2]{\mathbb{E}\,\(\,#1\,\big|\,#2\,\)\,}

\newcommand{\cVar}[2]{\Var\,\(\,#1\,\big|\,#2\,\)\,}

\newcommand{\sif}{$\sigma$-field}

\def\emailwww#1#2{\par\qquad {\tt #1}\par\qquad {\tt #2}\medskip}

%% file: intro.tex
Non-interacting particles are quanta of free quantum
fields.\footnote{%
 See \cite[Chaps 1,2]{BSZ}.}
Corresponding Hilbert spaces, known as Fock spaces,\footnote{%
 Or Fock-Cook representations, see \cite[Sect.~1.9, p.~61]{BSZ}.}
are direct sums of $ n $-particle spaces for $ n = 0,1,2,\dots $
Interacting particles are\footnote{%
 Or rather, should be; see \cite[Sect.~8.1]{BSZ}.}
described by nonlinear quantized fields that do not fit into Fock
spaces.

Fock spaces are also discussed by mathematicians in the context of
product systems, that is, continuous tensor products of Hilbert
spaces (see \cite{Ar} and references therein) and white noise
(see \cite{Ob, HKPS}). Non-classical noises found recently in
stochastic analysis (especially \cite{Wa}; see also \cite{TV}) are a
source of non-Fock product systems. It is natural to look at these
systems from the viewpoint of quantum field theory. I try to do so in
the present text by combining some ideas of Anatoly Vershik (the
initiator of the trend), Jonathan Warren, and mine.

Toy models considered here are quite poor in several
aspects. Especially, dynamics is neglected at all, and particles are
treated as utterly local in space (see Sect.~3.3). Nevertheless, we
may hope for some hints toward more useful rigorous non-Fock
constructions.

%% file: sect1.tex
\newcommand{\vac}{\psi_0}

\subsection{Lattice fermions}

The model can be described at once using notions of stochastic
analysis (see Sect.~2). However, for the beginning I prefer discrete
approximations. So, we assume for now that our ``physical space'' is a
one-dimensional lattice $ \eps \Z = \{ \dots, -2\eps, -\eps, 0, \eps,
2\eps, \dots \} $; we ascribe a two-dimensional Hilbert space $
H_{k\eps} $ to each lattice point $ k\eps $; we introduce on $
H_{k\eps} $ fermionic operators $ a_{k\eps} $ (annihilation), $
a^*_{k\eps} $ (creation), satisfying CAR (canonical anticommutation
relations\footnote{%
 Physicists are acquainted with CAR; mathematicians may look at
 \cite[Sect.~4.6]{Su}, \cite[Chap.~2]{BSZ}.}%
) $ a_{k\eps} a^*_{k\eps} + a^*_{k\eps} a_{k\eps} = 1 $, $ a_{k\eps}^2
= 0 $; we consider the tensor product\footnote{%
 In fact, only finite regions of the one-dimensional space will be
 used. The reader may restrict himself once and for all to $ k\eps \in
 [-1,+1] $, which makes $ H_{\eps\Z} $ finite-dimensional. Otherwise,
 see \cite[Sect.~4.5]{BSZ} (the direct product of grounded Hilbert
 spaces).}
$ H_{\eps\Z} $ of all $ H_{k\eps} $ containing the
vacuum\footnote{%
 It is a simplistic ``vacuum'', see Sect.~3.3.}
vector $ \vac $ such that $ a_{k\eps} \vac = 0 $ for all $ k $,
treating now all $ a_{k\eps} $ as operators on $ H_{\eps\Z} $, so that
$ a_{k\eps} a^*_{l\eps} + a^*_{l\eps} a_{k\eps} = 0 $ for $ k \ne l $,
and $ a_{k\eps} a_{l\eps} + a_{l\eps} a_{k\eps} = 0 $ for all $ k,l
$.\footnote{%
 To this end we identify $ a_{k\eps} \otimes a_{l\eps} $ with $
 a_{k\eps} a_{l\eps} $ whenever $ k < l $. The sign is chosen at
 will; see also \cite[Th.~3.2]{BSZ}.}
Of course, $ H_{\eps\Z} $ is just the usual (fermionic) Fock space;
a non-Fock space will appear later in the limit $ \eps \to 0 $, due to
a non-classical limiting procedure.
The total number of particles is the observable $ \bN^{\eps\Z} = \sum_k
a^*_{k\eps} a_{k\eps} $. Its eigenvalues are $ 0,1,2,\dots $ The
eigenspace $ H^{(n)}_{\eps\Z} $ corresponding to an eigenvalue $ n $
is the $ n $-particle space, spanned by vectors $ a^*_{k_1\eps} \ldots
a^*_{k_n\eps} \vac $ for $ k_1 < \dots < k_n $. The number of
particles on an interval $ [x,y) $ is the observable $ \bN^{\eps\Z}_{x,y} =
\sum_{k:x\le k\eps<y} a^*_{k\eps} a_{k\eps} $; clearly, $ \bN^{\eps\Z}_{x,y} +
\bN^{\eps\Z}_{y,z} = \bN^{\eps\Z}_{x,z} $ and $
\bN^{\eps\Z}_{-\infty,\infty} = \bN^{\eps\Z} $.

\subsection{Quantum states and classical random variables}

For each lattice point $ k\eps $ we introduce a variable $
\tau_{k\eps} $ with only two possible values $ \pm 1 $. Functions of
these variables will be used as symbols of state vectors, as follows:
\begin{equation}\label{1.2a}
\begin{matrix}
\text{function} & & & & \text{state vector} & & \\
1 & \text{(constant)} & & & \vac & \text{(vacuum)} & \\
\tau_{k\eps} & & & & a^*_{k\eps} \vac & & & \\
\tau_{k\eps} \tau_{l\eps} & & & & a^*_{k\eps} a^*_{l\eps} \vac & & (k<l)
\end{matrix}
\end{equation}
and so on; each factor $ \tau_{k\eps} $ gives rise to a particle at $
k\eps $. Extending the map by linearity we get a unitary operator from
the Hilbert space $ L_2 $ of all square integrable complex-valued
functions of variables $ \tau_{k\eps} $ onto the Hilbert space $
H_{\eps\Z} $ of all state vectors; $ L_2 \ni f \mapsto \psi_f \in
H_{\eps\Z} $.

Let $ f \in L_2 $, $ \| f \| = 1 $. Consider the probability $ \|
a_{k\eps} \psi_f \|^2 = \langle a^*_{k\eps} a_{k\eps} \rangle_{\psi_f}
$ of finding a particle at $ k\eps $. In terms of $ f $, it is the
so-called influence of $ \tau_{k\eps} $ on $ f $. For example, let $
\eps = 1 $ and $ f = f (\tau_1,\tau_2,\tau_3) $, then the influence of
$ \tau_1 $ on $ f $ is, by definition,
\[
\frac14 \sum_{a,b\in\{-1,+1\}} \bigg( \frac{ f(1,a,b) - f(-1,a,b) }{ 2
  } \bigg)^2 \, ;
\]
a probabilist writes it as the expectation $ \Ex $ of a conditional
variance $ \cVar{\cdot}{\cdot} $, treating $ \tau_{k\eps} $ as
independent random variables,\footnote{%
Just classical random variables on some probability space, with
(apriori) no relation to $ H_{\eps\Z} $.}
each with two equiprobable values: 
\begin{multline*}
\Ex \cVar{ f }{ \tau_2,\tau_3 } =
 \frac14 \sum_{s_2,s_3\in\{-1,+1\}}
 \bigg( \frac{
  f^2 (1,s_2,s_3) + f^2 (-1,s_2,s_3) }{ 2 } - \\
- \Big( \frac{ f (1,s_2,s_3) + f (-1,s_2,s_3) }{ 2
  } \Big)^2 \bigg) \, .
\end{multline*}
Similarly, the
probability of finding at least one particle within the two-point set
$ \{ k\eps, l\eps \} $ is the influence of $ \{ \tau_{k\eps},
\tau_{l\eps} \} $ on $ f $. For example, the influence of $ \{ \tau_1,
\tau_2 \} $ on $ f (\tau_1,\tau_2,\tau_3) $ is
\begin{multline*}
\Ex \cVar{ f }{ \tau_3 } = \\
 \frac12 \sum_{s_3\in\{-1,+1\}}
 \bigg( \frac{
  f^2 (1,1,s_3) + f^2 (1,-1,s_3) + f^2 (-1,1,s_3) + f^2 (-1,-1,s_3) }{ 4 } - \\
- \Big( \frac{ f (1,1,s_3) + f (1,-1,s_3) + f (-1,1,s_3) + f (-1,-1,s_3) }{ 4
  } \Big)^2 \bigg) \, .
\end{multline*}
In general, the probability of finding at least one particle on $
[x,y) $ is \newline
$ \Ex \cVar{ f }{ \tau_{\R\setminus[x,y)} } $, the
influence of $ \tau_{\R\setminus[x,y)} $ on $ f $; here $
\tau_{\R\setminus[x,y)} $ means the collection of $ \tau_{k\eps} $ for
all $ k $ such that $ k\eps \in \R\setminus[x,y) $, that is, $ k\eps <
x $ or $ k\eps \ge y $.

The corresponding quantum observable is a function of $ \bN^{\eps\Z}_{x,y}
$. Consider the projection onto the eigenspace of $ \bN^{\eps\Z}_{x,y} $
corresponding to the eigenvalue $ 0 $; it may be written as $
0^{\bN^{\eps\Z}_{x,y}} $ (it is means that $ 0^0 = 1 $, $ 0^1 = 0 $, $ 0^2 = 0
\dots $; note that $ 0^{\bN^{\eps\Z}_{x,y}} = \lim_{\la\to0} \la^{\bN^{\eps\Z}_{x,y}}
$). Its expectation $ \langle 0^{\bN^{\eps\Z}_{x,y}} \rangle_\psi $
(denoted also by $ \langle \psi | 0^{\bN^{\eps\Z}_{x,y}} | \psi \rangle $ or $
\( 0^{\bN^{\eps\Z}_{x,y}} \psi, \psi \) $) is the probability of finding no
particles on $ [x,y) $. So,
\begin{equation}\label{1.2b}
\langle 0^{\bN^{\eps\Z}_{x,y}} \rangle_{\psi_f} = 1 - \Ex \cVar{ f }{
\tau_{\R\setminus[x,y)} }
\end{equation}
whenever $ f \in L_2 $, $ \| f \| = 1 $.

\subsection{From a lattice to a continuum}

The standard limiting procedure is based on functions of the form
\[
f_{\phi,\eps} = \sqrt\eps \sum_{k\in\Z} \phi (k\eps) \tau_{k\eps}
\]
(as well as polynomials of such functions); here $ \phi : \R \to \C $
is a continuous compactly supported function (``test
function''). Clearly,
\[
\| f_{\phi,\eps} \|^2 = \eps \sum_k | \phi (k\eps) |^2
\xrightarrow[\eps\to0]{} \int_{-\infty}^{+\infty} | \phi (x) |^2 \, dx
= \| \phi \|^2 \, .
\]
The distribution of $ f_{\phi,\eps} $ tends (for $ \eps \to 0 $) to
the normal distribution $ N ( 0, \|\phi\|^2 ) $. This way, the array
of variables $ \tau_{k\eps} $ over the lattice $ \eps \Z $ turns (when
$ \eps \to 0 $) into the white noise $ \frac{d}{dx} \Br(x) $ over $ \R $;
here $ \Br $ is the Wiener process; $ \sqrt\eps \tau_{k\eps} $ turns
into $ d\Br(x) $. The function $ f_{\phi,\eps}
$ turns into the linear functional $ f_\phi = \int \phi(x) \, d\Br(x)
$. The lattice state vector $ \psi_{f_{\phi,\eps}} = \sqrt\eps \sum_k
\phi (k\eps) a^*_{k\eps} \vac $ turns into $
\psi_{f_\phi} = \int \phi (x) a^* (x) \vac \, dx $, the one-particle
state with the wave function $ \phi $; $ \frac1{\sqrt\eps} a_{k\eps} $
turns into $ a(x) $.\footnote{%
 Of course, $ a(x) $ is not an operator on $ H_\R^{\text{Fock}}
 $. Rather, $ a(\cdot) $ is a generalized function (Schwartz
 distribution) on $ \R $ taking on values in the space of linear
 operators on $ H_\R^{\text{Fock}} $.}
Similarly, for a function $ \phi $ of two variables, the quadratic
functional $ f_\phi = \iint\limits_{x<y} \phi(x,y) \, d\Br(x) d\Br(y) $
corresponds to the two-particle state $ \psi_{f_\phi} = \iint\limits_{x<y}
\phi(x,y) a^*(x) a^*(y) \vac \, dx dy $, and so on.
The number-of-particles operator corresponding to $ \sum_k a^*_{k\eps}
a_{k\eps} $ is naturally denoted by $ \bN = \int a^* (x) a(x) \, dx
$. Accordingly, $ \bN_{x,y} = \int_x^y a^* (z) a(z) \, dz $. Still, 
$ \bN_{x,y} + \bN_{y,z} = \bN_{x,z} $ and $ \bN_{-\infty,\infty} = \bN
$.
The standard
limiting procedure leads to the Fock space $ H_\R^{\text{Fock}} $, as
usual.

\subsection{Some specific random variables and corresponding quantum
 states}

We introduce functions $ W^{\eps\Z}_{k\eps} $ of variables $ \tau_{k\eps} $ as
follows:
\[
W^{\eps\Z}_0 = 0 \, ; \qquad W^{\eps\Z}_{(k+1)\eps} -
W^{\eps\Z}_{k\eps} = \sqrt\eps \tau_{k\eps} \quad \text{for } k \in \Z
\, .
\]
Random variables $ W^{\eps\Z}_{k\eps} $ form a random walk. Consider its
maximum over $ k\eps \in [-1,+1] $:
\[
W^{\eps\Z}_{X^{\eps\Z}} = \max_{k\in\Z\cap[-1/\eps,1/\eps]}
W^{\eps\Z}_{k\eps} \, ; \qquad X_{\eps\Z} =
\argmax_{x\in\eps\Z\cap[-1,+1]} W^{\eps\Z}_x \, .
\]
For almost all $ \tau_{k\eps} $ the maximum is reached only once, thus
$ X^{\eps\Z} $ is a well-defined random variable (a function of random
variables $ \tau_{k\eps} $) whose values belong to the finite piece $
\eps\Z \cap [-1,1] $ of the lattice. Also $ W^{\eps\Z}_{X^{\eps\Z}} $ is a random
variable. Both determine state vectors $ \psi_{X^{\eps\Z}} $, $ \psi_{W^{\eps\Z}_{X^{\eps\Z}}}
$. These are quite complicated linear combinations of various
multi-partical states. In order to calculate them explicitly, one
should evaluate Fourier-Walsh coefficients by averaging $
\tau_{k_1\eps} \dots \tau_{k_n\eps} X^{\eps\Z} $ and $ \tau_{k_1\eps} \dots
\tau_{k_n\eps} W^{\eps\Z}_{X^{\eps\Z}} $; fortunately we do not need it. The standard
limiting procedure $ \eps \to 0 $ gives the Wiener process $ \Br(x) $,
and $ \Br(X) = 
\max_{x\in[-1,1]} \Br(x) $, $ X = \argmax_{[-1,1]} \Br(\cdot) $ (also
reached only once, almost sure). Fourier-Walsh expansion of $
X^{\eps\Z} $ becomes Wiener chaos expansion of $ X $ into It\^o's
multiple stochastic integrals,
\begin{equation}\label{1.4a}
X = \hat X_0 + \int \hat X_1 (x) \, d\Br(x) + \iint\limits_{x<y} \hat
X_2 (x,y) \, d\Br(x) d\Br(y) + \dots
\end{equation}
The function $ \hat X_k (x_1,\dots,x_k) $ for $ -1 < x_1 < \dots < x_k
< 1 $ is the wave function of the $ k $-particle component of $ \psi_X
$. The squared norm of $ \psi_X $ is the average of $ X^2 $, equal to
$ 1/2 $ since $ \arcsin X $ is known to be distributed uniformly on $
(-\pi/2, \pi/2) $. Still, it happens in the Fock space.

Trying to escape the Fock space, we consider such functions as $ \exp
(i\la X^{\eps\Z}) $ and $ \exp ( i \la W^{\eps\Z}_{X^{\eps\Z}} ) $; here $ \la $ is a large
parameter (a kind of cutoff parameter), it will ultimately tend to $
\infty $. Clearly, both functions become senseless for $ \la = \infty
$. However, it is instructive to compare the behavior of $ \psi_{\exp
(i\la X^{\eps\Z})} $ and $ \psi_{\exp (i\la W^{\eps\Z}_{X^{\eps\Z}})} $ for large $ \la $.

In order to estimate (via \eqref{1.2b}) the density of particles in
the state $ \psi_{\exp (i\la W^{\eps\Z}_{X^{\eps\Z}})} $, consider the
influence of a single variable $ \tau_{k\eps} $ on $ \exp (i\la
W^{\eps\Z}_{X^{\eps\Z}}) $. It is an average over paths of the
walk. Assume $ k\eps \in (0,1) $. A path
that reaches its maximum on $ (-1,k\eps) $ gives a small (in most
cases, just $ 0 $) contribution to the influence. A path that reaches
its maximum on $ (k\eps,1) $ contributes (in most cases) roughly $
(\la\sqrt\eps)^2 $, assuming that $ \eps \ll 1/\la^2 $. In the limit $
\eps \to 0 $ the same argument, applied to the Wiener process, shows
that the density of the particles on $ (-1,+1) $ is roughly $ \la^2
$; too much for taking $ \la\to\infty $.

Consider now the density of particles in the other state, $ \psi_{\exp
(i\la X^{\eps\Z})} $. The influence of $ \tau_{k\eps} $ on $ \exp (i\la X^{\eps\Z}) $ is
an average over paths. Assume again $ k\eps \in (0,1) $, and consider
a path that reaches its maximum on $ (k\eps,1) $, not too close to $
k\eps $. Its contribution to the influence is in most cases exactly $
0 $, irrespective of $ \la $. A non-zero contribution appears only
when the maximum on $ (-1,k\eps) $ is $ \sqrt\eps $-close to the
maximum on $ (k\eps,1) $. In that case, $ \tau_{k\eps} $ influences $
X^{\eps\Z} $ by causing a jump of a size roughly $ 1 $. Also $ \exp
(i\la X^{\eps\Z}) $ makes a jump of a size roughly $ 1 $ irrespective of $
\la $. The probability of a particle at a given point of $ \eps\Z $ is
$ \sim \sqrt\eps $ for large $ \la $. The same argument, applied on
the continuum ($ \eps = 0 $) shows that the probability of at least
one particle on $ (x,x+\Delta x) $ is $ \sim \sqrt{\Delta x} $ for
large $ \la $. This is why the limit $ \la\to\infty $ is of interest
for such a state.

Of course, the expression $ \exp ( i \infty X ) $
is not a well-defined function over the Wiener process, because of its
infinite sensitivity to small changes of $ \Br(\cdot) $. However, the
sensitivity is concentrated near the maximizer. Accordingly, particles
accumulate to a single point (though it does not follow from the
simple argument of the preceding paragraph).

%% file: sect2.tex
\newcommand{\LWi}{L_2^{\text{Wiener}}}
\newcommand{\Lsp}{L_2^{\text{splitting}}}
\newcommand{\HFo}{H^{\text{Fock}}}

\subsection{Beyond the white noise}

\begin{figure}[tb]
\begin{center}
\setlength{\unitlength}{1cm}
\begin{picture}(12,2.6)
\put(-0.7,-0.4){\includegraphics{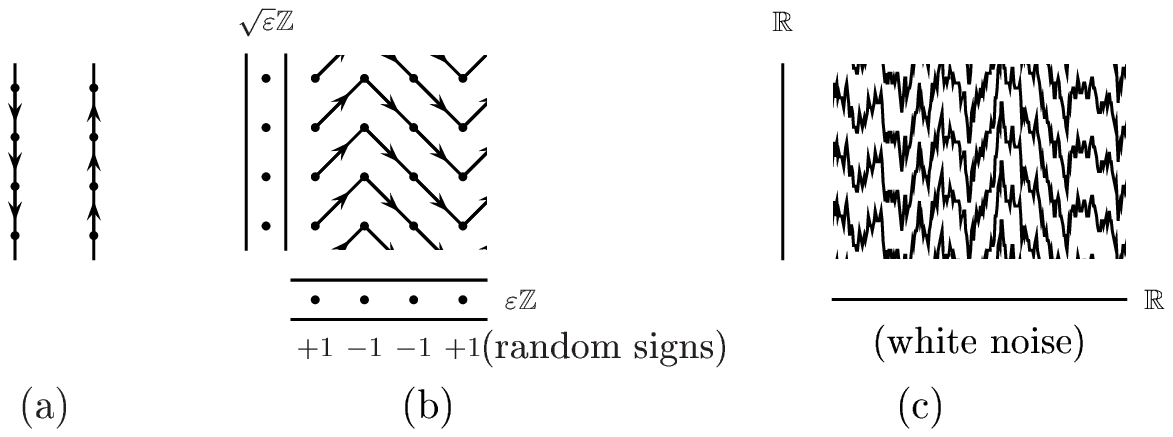}}
\end{picture}
\caption[]{\label{fig1}\small
Two simple transformations of the lattice $ \sqrt\eps\Z $ (a),
alternating at random, form the simple lattice stochastic flow (b),
consisting of parallel copies of the random walk. Its scaling limit ($
\eps \to 0 $) is the simple stochastic flow (c), consisting of
parallel copies of the Wiener process.}
\end{center}
\end{figure}

The Wiener process is the scaling limit of the random walk, which may
be explained as follows. Consider the lattice $ \eps\Z = \{ \dots, -2\eps, -\eps, 0, \eps,
2\eps, \dots \} $, another lattice $ \sqrt\eps\Z = \{ \dots,
-2\sqrt\eps, -\sqrt\eps, 0, \sqrt\eps, 2\sqrt\eps, \dots \} $, and
maps $ U_{x,y} : \sqrt\eps\Z \to \sqrt\eps\Z $ defined for $ x,y \in
\eps\Z $, $ x < y $:
\[
U_{k\eps,l\eps} (\sqrt\eps m) = \sqrt\eps ( m + \tau_{k\eps} +
\tau_{(k+1)\eps} + \dots + \tau_{(l-1)\eps} ) \, ;
\]
here, as before, $ (\tau_{k\eps})_{k\in\Z} $ is an array of
independent random signs. Note that $ U_{y,z} \( U_{x,y} (u) \) =
U_{x,z} (u) $, that is, $ U_{y,z} \circ U_{x,y} = U_{x,z} $ for $ x <
y < z $; all $ U_{x,y} $ are compositions of maps $ U_{x,x+\eps} $ ($
x \in \eps\Z $), these being independent; each $ U_{x,x+\eps} $ is
either $ u \mapsto u+\sqrt\eps $ or $ u \mapsto u-\sqrt\eps $. The
lattice stochastic flow (Fig.~\ref{fig1}b) results from random
alternating the two simple transformations of $ \sqrt\eps\Z $
(Fig.~\ref{fig1}a). The standard limiting procedure for $ \eps \to 0 $
leads to $ U_{x,y} (u) = u + \Br(y) - \Br(x) $, $ u \in \R $, $ x,y
\in \R $, $ x \le y $ (Fig.~\ref{fig1}c); as before, $ \Br(\cdot) $ is
the Wiener process.

\begin{figure}[tb]
\begin{center}
\setlength{\unitlength}{1cm}
\begin{picture}(12,5)
\put(-0.7,-0.4){\includegraphics{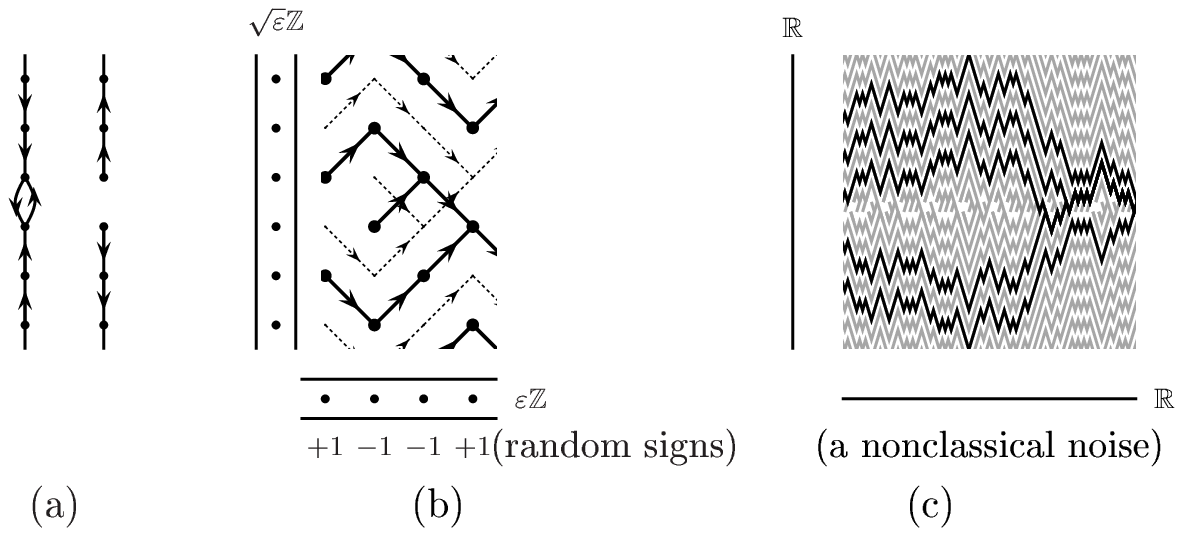}}
\end{picture}
\caption[]{\label{fig2}\small
Another pair of transformations of $ \sqrt\eps\Z $ (a); their random
alternation form another lattice stochastic flow (b). Its scaling
limit ($ \eps \to 0 $) is the splitting stochastic flow (c).}
\end{center}
\end{figure}

Now, breaking homogeneity on the $ u $ axis, we consider another pair
of transformations $ \sqrt\eps\Z \to \sqrt\eps\Z $
(Fig.~\ref{fig2}a). Alternating them at random we get another lattice
stochastic flow\footnote{%
The two-dimensional lattice $ (\sqrt\eps\Z) \times (\eps\Z) $
decomposes into two sublattices (even and odd), closed under the
flow. In order to get a scaling limit we restrict ourselves to one of
the two sublattices.}
(Fig.~\ref{fig2}b) that has its scaling limit (Fig.~\ref{fig2}c);
we'll call it \emph{the splitting flow.}

For large $ u $, near $ +\infty $, the splitting flow does not differ
from the simple flow:
\begin{equation}\label{2.1a}
U_{x,y} (u) = u + \Br(y) - \Br(x) \qquad \text{for } u > \Br(x) -
\min_{[x,y]} \Br(\cdot) \, ;
\end{equation}
$ \Br(\cdot) $ is still the Wiener process. Near $ -\infty $ the
situation is symmetric. Near $ 0 $ we have
\[
U_{x,y} (u) = \pm \( \Br(y) - \min_{[x,y]} \Br(\cdot) \) \qquad
\text{for } |u| < \Br(x) - \min_{[x,y]} \Br(\cdot) \, .
\]
Thus, for given $ x,y \in \R $, $ x < y $, the map $ U_{x,y} : \R \to
\R $ is chosen at random from a two-dimensional set $ G $ of
maps,\footnote{%
 The set $ G $ is a semigroup, and a (two-dimensional) topological
 space. However, it is not a group (maps are not invertible), nor even
 a topological semigroup (maps are discontinuous, and the composition
 is a discontinuous operation on the semigroup).}
determined by two parameters, $ \Br(x) - \min_{[x,y]} \Br(\cdot) $ and
$ \pm \( \Br(y) - \min_{[x,y]} \Br(\cdot) \) $. The first parameter is
uniquely determined by $ \Br(\cdot) $, but the second is not; its sign
is a random variable independent of $ \Br(\cdot) $. On the lattice, the
sign is determined by the parity (evenness) of the number $ k $ of the
point $ k\eps \in [x,y] $ that minimizes $ \Br^{\eps\Z}_{k\eps} $. In terms of
Sect.~1.4 the sign is just $ \exp (i\la X) $ for $ \la = \pi / \eps $ (though,
minimum is used now, rather than maximum). For $ \eps \to 0 $ it turns
into the ill-formed expression $ \exp (i\infty X) $. However, the sign
itself is well-defined, and may be used as a well-formed substitute
for the ill-formed expression!

The random map $ U_{x,y} $, that is, the family of random variables $
\{ U_{x,y} (u) : u \in \R \} $ (for given $ x,y $) reduces to
independent components: a single random sign $\, V_{x,y} = \sgn U_{x,y}
(0) $, and increments $ \{ \Br(b) - \Br(a) : x<a<b<y \} $ of the
Wiener process. The whole splitting flow $ \{ U_{x,y} (u) : u \in \R,
-\infty < x < y < +\infty \} $ boils down to random signs $ \{ V_{x,y} :
x<y \} $ and Wiener process increments $ \{ \Br(b) - \Br(a) : a<b \}
$. However, these are dependent;
\begin{equation}\label{2.1b}
\begin{aligned}
\cE{ V_{x,y} }{ \Br(\cdot) } &= 0 \, , \\
\cE{ V_{x_1,y_1} V_{x_2,y_2} }{ \Br(\cdot) } &= \begin{cases} 1 &
  \text{if $ \argmin_{[x_1,y_1]} \Br(\cdot) = \argmin_{[x_2,y_2]}
        \Br(\cdot) $}, \\ 0 & \text{otherwise}. \end{cases}
\end{aligned}
\end{equation}
Given $ \Br(\cdot) $, random signs $ V_{x,y} $ depend on $ x,y $ only
via $ \argmin_{[x,y]} \Br(\cdot) $,
\[
V_{x,y} = V_{\argmin_{[x,y]} \Br(\cdot)} \, .
\]
Random signs $ V_x $ are ascribed to all points $ x $ of local minima
of $ \Br(\cdot) $. Such points are a dense countable set. In fact,
\[
V_x = \lim_{\eps\to0} \sgn U_{x-\eps,x+\eps} (0) \, ;
\]
the limit exists if (and only if) $ \Br(\cdot) $ has a local minimum
at $ x $.

\subsection{From random processes to Hilbert spaces}

The Wiener process $ \( \Br(x) \)_{-\infty<x<\infty} $ is a Brownian
motion (in the mathematical sense rather than physical) in the
additive group $ \R $ of real numbers. The splitting flow $ \(
U_{x,y}(\cdot) \)_{-\infty<x<y<\infty} $ is a Brownian motion in the
two-dimensional non-commutatite semigroup $ G $. For the Wiener
process, only increments are relevant, $ \( \Br(y) - \Br(x)
\)_{-\infty<x<y<\infty} $. For the splitting flow we cannot write $
U_{x,y} = U_{0,y} \circ U^{-1}_{0,x} $ since $ U_{0,x} $ is not
invertible; this is why we use $ U_{x,y} $ rather than $ U(x) =
U_{0,x} $.

All square integrable measurable functionals of the WIener process are
a separable Hilbert space $ \LWi (-\infty,+\infty) $. Given $ (x,y)
\subset \R $ we may consider functionals of increments $ \Br(b) -
\Br(a) $ for all $ (a,b) \subset (x,y) $; such functionals form $ \LWi
(x,y) $. Independence of increments gives
\[
\LWi (x,y) \otimes \LWi (y,z) = \LWi (x,z) \, .
\]

All square integrable measurable functionals of the splitting flow are
a separable Hilbert space $ \Lsp (-\infty,+\infty) $. Given $ (x,y)
\subset \R $ we may consider functionals of $ U_{a,b} (\cdot) $ for
all $ (a,b) \subset (x,y) $; such functionals form $ \Lsp (x,y)
$. Independence\footnote{%
 I mean independence between two families of random variables, $ \{
 U_{a,b} (u) : u \in \R, (a,b) \subset (x,y) \} $ and $ \{ U_{a,b} (u)
 : u \in \R, (a,b) \subset (y,z) \} $.}
gives
\[
\Lsp (x,y) \otimes \Lsp (y,z) = \Lsp (x,z) \, .
\]
The Wiener process is naturally embedded into the splitting flow by
\eqref{2.1a}, thus,
\[
\LWi (x,y) \subset \Lsp (x,y) \, .
\]
The space $ \LWi (-\infty,+\infty) $ may be identified\footnote{%
 Recall Sect.~1.3.}
with the Fock space,
\[
\HFo_{\R} = \LWi (-\infty,+\infty) \, ,
\]
over the space $ H_{\R}^{(1)} $ of all one-particle state vectors,
identified with linear functionals $ \int \phi(x) \, dW(x) $. We have
also local spaces\footnote{%
 There are two equally canonical ways of identifying $ \HFo_{(x,y)}
 \otimes \HFo_{(y,z)} $ with $ \HFo_{(x,z)} $. Namely, for $ x_1 \in
 (x,y) $, $ x_2 \in (y,z) $ we may identify $ a(x_1) \otimes a(x_2) $
 with $ a(x_1) a(x_2) $ or alternatively with $ a(x_2) a(x_1) = -
 a(x_1) a(x_2) $. See also \cite[Th.~3.2]{BSZ}.}
\begin{gather*}
\HFo_{(x,y)} = \LWi (x,y) \, , \\
\HFo_{(x,y)} \otimes \HFo_{(y,z)} = \HFo_{(x,z)} \, , \\
H_{(x,y)}^{(1)} \oplus H_{(y,z)}^{(1)} = H_{(x,z)}^{(1)} \, .
\end{gather*}
Now we \emph{define} non-Fock spaces of our model by
\begin{gather*}
H_{\R} = \Lsp (-\infty,+\infty) \, , \\
H_{(x,y)} = \Lsp (x,y) \, ,
\end{gather*}
which gives
\begin{equation}\label{2.2a}
\begin{gathered}
H_{(x,y)} \otimes H_{(y,z)} = H_{(x,z)} \, , \\
\HFo_{(x,y)} \subset H_{(x,y)} \, .
\end{gathered}
\end{equation}

\subsection{Infinitely many particles}

How to count particles in a non-Fock state $ \psi \in H_{\R} \, $? The
number-of-particles operator $ \bN $ is defined on $ \HFo_{\R} $
only. Turn to $ 0^\bN $. This observable indicates the event ``no
particles''; it is the one-dimensional projection $ |\vac\rangle
\langle\vac| $ onto the vacuum $ \vac $. It is natural to postulate
that in the non-Fock model the vacuum is still the only state with no
particles.\footnote{%
 Thus, ``to contain particles'' means ``to be orthogonal to $ \vac
 $''. True, it is not a deep definition of ``particles''. We'll see
 soon that our ``particles'' can be localized in space (too much for
 physical particles, see Sect.~3.3). Further they will get an electric
 charge, see Sect.~3.2.}
That is, we extend $ 0^\bN $ from $ \HFo_\R $ onto $ H_\R $ by $ 0^\bN
= |\vac\rangle \langle\vac| $. In probabilistic terms $ \vac $ is just
$ 1 $ (recall \eqref{1.2a}), and the corresponding one-dimensional
projection is just $ \Ex $, the operator of mathematical expectation;
$ 0^\bN f = (\Ex f) \cdot \vac $, that is, $ 0^\bN f = \Ex f $ for $ f
\in \Lsp (-\infty,+\infty) $.

We localize the argument to $ (x,y) \subset \R $. Namely, we postulate
that in the (local) non-Fock space $ H_{(x,y)} $ the only state with
no particles is the (local) vacuum state $ \vac^{(x,y)} $ (belonging
to $ \HFo_{(x,y)} $) identified with the constant random variable $ 1
$. That is, we have
\begin{gather*}
\vac^{(x,y)} \otimes \vac^{(y,z)} = \vac^{(x,z)} \, , \\
\vac = \vac^{(-\infty,+\infty)} \, ,
\end{gather*}
and we extend $ 0^{\bN_{x,y}} $ from $ \HFo_{(x,y)} $ onto $ H_{(x,y)}
$ by
\[
0^{\bN_{x,y}} = | \vac^{(x,y)} \rangle \langle \vac^{(x,y)} | \, ,
\]
the one-dimensional projection onto the local vacuum. Alternatively,
we may treat $ 0^{\bN_{x,y}} $ as an operator on the whole space:
\begin{align*}
H_\R &= H_{(-\infty,x)} \otimes H_{(x,y)} \otimes H_{(y,+\infty)} \, ,
  \\
0^{\bN_{x,y}} &= 1 \otimes \( | \vac^{(x,y)} \rangle \langle
  \vac^{(x,y)} | \) \otimes 1 \, ,
\end{align*}
the projection onto the (infinite-dimensional) space $ H_{(-\infty,x)}
\otimes \vac^{(x,y)} \otimes H_{(y,+\infty)} $. In probabilistic
terms, the space consists of all functionals of $ \{ U_{x_1,x_2} (u) : u
\in \R, (x_1,x_2) \subset (-\infty,x) \} \cup \{ U_{x_1,x_2} (u) : u \in \R,
(x_1,x_2) \subset (y,+\infty) \} $, and the projection is the conditional
expectation,
\[
0^{\bN_{x,y}} f = \cE{ f }{ \F_{(-\infty,x)\cup(y,+\infty)} } \, ;
\]
$ \F_E $ stands for the \sif\ (on the probability space of the
splitting flow) generated by the set $ \{ U_{a,b} (u) : u \in \R,
(a,b) \subset E \} $ of random variables, whenever $ E \subset \R $ is
an interval or a union of finitely many intervals.

We have commuting observables $ 0^{\bN_{x,y}} $ on $ H_\R $,
satisfying
\begin{equation}\label{2.3a}
0^{\bN_{x,y}} 0^{\bN_{y,z}} = 0^{\bN_{x,z}} \, .
\end{equation}
We may consider the corresponding joint quantum measurement. Its
outcome assigns values ($ 0 $ or $ 1 $) to all $ \bN_{x,y} $
respecting \eqref{2.3a}, and is interpreted as a configuration of
particles in the
(one-dimensional) space. If $ \psi \in \HFo_\R $ then the outcome is
always a finite set. For $ \psi \in H_\R $ the outcome can be
infinite. In fact, if $ \psi $ is orthogonal to the Fock space (that
is, $ \psi \in H_\R \ominus \HFo_\R $), then the outcome is always an
infinite set.

More formally, every state vector $ \psi \in H_\R $, $ \| \psi \| = 1
$, determines a probability measure $ \mu_\psi $ on the space of
compact subsets of $ \R $, defined by
\[
\forall E \qquad \langle 0^{\bN_E} \rangle_\psi = \mu_\psi \{ C : C
\cap E = \emptyset \} \, ;
\]
here $ E $ runs over finite unions of intervals; $ 0^{\bN_E} $ is
defined naturally:
\[
0^{\bN_E} = 0^{\bN_{x_1,y_1}} \cdot \ldots \cdot 0^{\bN_{x_n,y_n}}
\qquad \text{for } E = (x_1,y_1) \cup \dots \cup (x_n,y_n) \, ;
\]
in fact, \eqref{2.3a} is a special case of $ 0^{\bN_{E_1\cup E_2}} =
0^{\bN_{E_1}} \cdot 0^{\bN_{E_2}} $. See \cite[Sect.~2]{unitary} and
also \cite{Ts:Fe} for a general theory of $ \mu_\psi $.

If $ \psi = \psi_0 $ then $ \mu_\psi $ is concentrated on $ \{
\emptyset \} $. That is, if we measure the configuration of particles
in vacuum, we always get the empty set $ \emptyset $.

If $ \psi \in \HFo_\R $ then $ \mu_\psi $ is concentrated on (the set
of) finite sets. That is, if we measure the configuration of particles
in a Fock state, we get a random finite set.

If $ \psi \in H_\R \ominus \HFo_\R $ then $ \mu_\psi $ is concentrated
on (the set of) infinite sets. That is, if we measure the configuration
of particles in a non-Fock state (orthogonal to the Fock space), we
get a random infinite set. The general theory
\cite{unitary} states that $ \mu_\psi \{ C : C \ni x \} = 0 $ for
every $ x $ (and $ \psi $). It follows that $ \mu_\psi $ is
concentrated on (the set of) sets $ C $ of zero Lebesgue measure. For
some models,
$ C $ is always of cardinality continuum.

Return to our particular model, $ H_\R = \Lsp $. Recall that $ \Lsp $
is generated by $ \LWi $ and random signs $ V_{x,y} $. The simplest
example of $ \psi \in H_\R \ominus \HFo_\R $ is $ \psi = V_{0,1}
$. The corresponding $ \mu_\psi $ is described, rather explicitly, by
Warren \cite{Wa}. It is concentrated on (the set of) infinite,
countable sets $ C $
with a single accumulation point (which was roughly explained here, in
Sect.~1.4). That is, if we measure the configuration of particles in
such a state, we always get a random countable set with a single,
random, accumulation point.

%% file: sect3.tex
\subsection{Three-dimensional space}

Recall the idea explained in Sect.~1.4: the maximizer $ X =
\argmax_{[-1,+1]} W(\cdot) $ of the Wiener process is rather
insensitive to changes of $ W(\cdot) $ outside of a neighborhood of $
X $. The set of local maxima (or minima) of $ W(\cdot) $ is a (random)
dense countable set. And, as explained in Sect.~2.1, the insensitivity
emerges from locality of the notion of a local maximum.

Turning from $ \R $ to the 3-dimensional space $ \R^3 $, we turn from
$ W(\cdot) $ to the white noise over $ \R^3 $; its sample function is
a generalized function (Schwartz distribution) and gives a number,
being integrated over some 3-dimensional domain. The integral is a
random variable distributed normally with zero mean; its variance is
equal to the volume of the domain.

A straightforward attempt to generalize $ \argmax W(\cdot) $ for $
\R^3 $ meets many difficulties. However, we need only \emph{some}
dense countable set of (random) points, determined by a local property
of a white noise sample function. I am unable to find such a property
among natural, well-known properties. Instead, I am able to invent
such a property.

Consider the integral $ S (x,r) $ of the 3-dimensional white noise
over the 3-dimensional ball $ B(x,r) \subset \R^3 $ of radius $ r \in
(0,\infty) $, centered at $ x \in \R^3 $. Given $ x_0 $ and $ r_0 $,
we consider all $ x_1, r_1 $ such that $ B(x_1,r_1) \subset B(x_0,r_0)
$ and $ r_1 \le r_0/2 $. We choose $ x_1, r_1 $ maximizing $ |
S(x_1,r_1) | $; the maximum is reached only once (with probability
1). Having $ x_1, r_1 $ we choose $ x_2, r_2 $ in the same way: $
B(x_2,r_2) \subset B(x_1,r_1) $, $ r_2 \le r_1/2 $, and $ (x_2,r_2) =
\argmax_{(x_2,r_2)} |S(x_2,r_2)| $. And so on. The sequence $ (x_n) $
converges to a point: $ x_n \to X_{x_0,r_0} $. We have a family $
\(X_{x_0,r_0}\)_{x_0\in\R^3,r_0\in(0,\infty)} $ of $ \R^3 $-valued
random variables; they are functionals of the 3-dimensional white
noise. A small change of $ (x_0,r_0) $ does not influence $
X_{x_0,r_0} $, therefore the random set $ \{ X_{x_0,r_0} : x_0\in\R^3,
r_0\in(0,\infty) \} $ is (at most) countable. The set is dense, since
$ r_0 $ may be arbitrarily small.

Now we can mimick Warren's construction as follows. We introduce
random signs $ V_{x,r} $ such that (recall \eqref{2.1b})
\[
\begin{aligned}
\cE{ V_{x,r} }{ S(\cdot,\cdot) } &= 0 \, , \\
\cE{ V_{x_1,r_1} V_{x_2,r_2} }{ S(\cdot,\cdot) } &= \begin{cases} 1 &
  \text{if $ X_{x_1,r_1} = X_{x_2,r_2} $}, \\ 0 &
\text{otherwise}. \end{cases}
\end{aligned}
\]
In other words, we ascribe new random signs $ V_x $ to all points of
the dense countable random set.

Every smoothly bounded domain $ D \subset \R^3 $ determines its
Hilbert space $ \HFo_D $, generated by $ S (x,r) $ for all $ (x,r) $
such that $ B (x,r) \subset D $. Also, it determines a larger space $
H_D $ generated by $ S(x,r) $ and $ V_{x,r} $ for all $ (x,r) $ such
that $ B(x,r) \subset D $. Similarly to \eqref{2.2a},
\[
\begin{gathered}
\HFo_D \subset H_D \, , \\
\HFo_{D_1} \otimes \HFo_{D_2} = \HFo_D \, , \\
H_{D_1} \otimes H_{D_2} = H_D
\end{gathered}
\]
whenever $ D $ is split into two domains $ D_1, D_2 $ by a smooth
surface.\footnote{%
 Fock spaces $ \HFo_D $ can be defined also for non-smooth domains $ D $
 without loss of their natural properties, but non-Fock spaces $ H_D $
 cannot; see \cite{Ts:Fe}.}

The random sign $ V_{x,r} $ is identified with a state vector $
\psi_{x,r} \in H_D \ominus \HFo_D $, where $ D = B (x,r) $. If we
measure the configuration of particles in such a state, we always get
a random countable subset of $ D $ with a single, random, accumulation
point.

\subsection{Electric charge}

Charged particles are quanta of complex fields. For the Fock model it
may be interpreted via the complex-valued (rather than real-valued)
white noise. In one-dimensional space, every square-integrable
functional of the complex-valued Wiener process $ Z (x) = W_1(x) +
iW_2(x) $ (where $ W_1, W_2 $ are independent real-valued Wiener
processes) has its Wiener chaos expansion similar to \eqref{1.4a}, but
instead of $ dW(x), dW(x) dW(y) $ and so on, we get $ dZ(x),
d\overline Z(x), dZ(x) dZ(y), dZ(x) d\overline Z(y), d\overline Z(x)
dZ(y), d\overline Z(x) d\overline Z(y) $ and so on; here $ \overline
Z(x) $ is the complex conjugate to $ Z(x) $. The term with $ dZ(x) $
describes a single particle with the charge $ +1 $; the charge $ -1 $
is described by $ d\overline Z(x) $; $ dZ(x) d\overline Z(y) $
describes two particles, one of positive charge (at $ x $), the other
of negative charge (at $ y $); and so on. The sector of (total) charge
$ 0 $ consists of functionals invariant under the global commutative
gauge group $ Z(\cdot) \mapsto e^{i\phi} Z(\cdot) $. The corresponding
one-parameter unitary group on $ H_\R $ has its generator, the
observable of total electric charge $ Q $. Localization to $ H_{(x,y)}
$ gives $ Q_{(x,y)} $, the electric charge on $ (x,y) $.

As far as I know, there is no interesting generalization to $ Z(\cdot)
$ of the idea of Sect.~1.4, based on $ \argmax_{[x,y]} W(\cdot)
$.\footnote{%
 We cannot place random signs at local maxima of $ \Re Z(\cdot) $,
 since $ \Re \( e^{i\phi} Z(\cdot) \) $ has quite different local
 minima; the union over all $ \phi $ is uncountable. Local minima of $
 |Z(\cdot)| $, that is, of $ | \int_0^x dZ(\cdot) | $, lead to a
 theory treating the origin as a special point in the one-dimensional
 space.}
However, the idea of Sect.~3.1 can be used easily. Still, $ |
S(x_1,r_1) | $ is maximized, but now $ S(x_1,r_1) $ is a complex
number. The maximizer is invariant under $ Z(\cdot) \mapsto e^{i\phi}
Z(\cdot) $. We extend the gauge group from $ \HFo $ to $ H $ by
postulating that random signs $ V_{x,r} $ are gauge invariant. Thus,
$ V_{x,r} $ is a state vector belonging to the zero-charge sector.

Local gauge transformations $ Z(x) \mapsto e^{i\phi(x)} Z(x) $
determine the (infinite-dimensional, unitary) gauge group on $ H_\R $
(still, $ V_{x,r} $ are gauge-invariant by definition). The group
commutes with the observables $ 0^{\bN_{x,y}} $, which means that we
can measure both the spatial configuration of particles and
(simultaneously) the spatial distribution of electric charge. An
outcome of such a measurement in the state $ V_{x,r} $ is always a
countable set of particles, each having a charge $ \pm 1 $. As was
said, the set has a single accumulation point. The charge in a
neighborhood of the accumulation point is determined by the fact that
the total charge is equal to zero.

\subsection{Dirac sea and dynamics}

Non-interacting electrons (and positrons) are quanta of the free Dirac
quantum field,\footnote{%
 See \cite[Sect.~7.2]{Su}, \cite[Sect.~6.5]{BSZ}.}
much more singular than our toy models. The distribution of the
electric charge can be treated as a generalized function (Schwartz
distribution) on 4-dimensional space-time rather than 3-dimensional
space; otherwise fluctuations become infinite.\footnote{
 See \cite[Sect.~7]{Th}.}
In terms of the lattice approximation of Sect.~1.1 we have, roughly
speaking,\footnote{%
 The space should be 3-dimensional; a spinor index should be added to
 the spatial index $ k $; the infinite product is ill-defined.}
\[
\psi^{\text{Dirac}}_0 = \bigg( \prod_n \bigg( \sum_k \phi_n (k\eps)
a^*_{k\eps} \bigg) \bigg) \psi_0 \, ;
\]
here $ \psi_0 $ is the simplistic vacuum of Sect.~1.1; $
\psi^{\text{Dirac}}_0 $ is the vacuum of the free Dirac quantum field;
and functions $ \phi_n (\cdot) $ on the lattice $ \eps\Z $ form an
orthonormal basis of a special subspace of $ L_2 (\eps\Z) $, so-called
negative energy subspace. It is infinite-dimensional, moreover, it is
a half of $ L_2 (\eps\Z) $ in the sense that $ \langle a_{k\eps}^*
a_{k\eps} \rangle_{\psi^{\text{Dirac}}_0} = 1/2 $ for all $ k $; each
lattice point is occupied with probability $ 1/2 $ (these events being
correlated).

According to Dirac, the vacuum is such a state: all negative-energy
states are occupied, while all positive-energy states are
free. Positrons are holes in a sea of negative electrons. See
\cite[Sect.~6.5]{BSZ} for a rigorous definition of the relevant space
that contains $ \psi^{\text{Dirac}}_0 $ (and does not contain $ \psi_0
$). Particles (electrons and positrons) are inherently delocalized,
see \cite[Sect.~1]{Th}.

What about ``filling in the Dirac sea'' for non-Fock models considered
here? For now it remains unclear. Maybe the ``spatial'' approach to
product systems in quantum field theory is too naive; for a different,
4-dimensional approach see \cite[Sect.~1]{Ar}.

%% file: main.bbl
\begin{thebibliography}{99}

\bibitem{Ar} W.~Arveson,
``$ E_0 $-semigroups in quantum field theory'',
Proc. Sympos. Pure Math. \textbf{59} (1996), 1--26.

\bibitem{BSZ}
J.C.~Baez, I.E.~Segal, Z.~Zhou,
``Introduction to algebraic and constructive quantum field theory'',
Princeton Univ.\ Press 1992.

\bibitem{HKPS} T.~Hida, H-H.~Kuo, J.~Potthoff, L.~Streit,
``White noise: an infinite dimensional calculus'',
Kluwer Academic Publ.\ 1993.

\bibitem{Ob} N.~Obata,
``White noise calculus and Fock space'',
Lecture Notes in Mathematics (Springer) \textbf{1577} (1994).

\bibitem{Su} A.~Sudbery,
``Quantum mechanics and the particles of nature: An outline for
mathematicians'',
Cambridge Univ. Press, 1986.

\bibitem{Th} W.E.~Thirring,
``Principles of quantum electrodynamics'',
Acad. Press, 1958.

\bibitem{Ts:Fe} B.~Tsirelson, ``Noise sensitivity on continuous
products: an answer to an old question of J.~Feldman'',
math.PR/9907011.

\bibitem{unitary} B.~Tsirelson, ``Unitary Brownian motions are
linearizable'', math.PR/9806112.

\bibitem{TV} B.S.~Tsirelson, A.M.~Vershik,
``Examples of nonlinear continuous tensor products of measure spaces
and non-Fock factorizations'',
Reviews in Mathematical Physics \textbf{10}:1 (1998), 81--145.

\bibitem{Wa} J.~Warren, ``Splitting: Tanaka's SDE revisited'',
math.PR/9911115.

\end{thebibliography}
